\documentclass[aps,prl,amsfonts,amsmath,nofootinbib,showpacs,preprint,tightenlines,longbibliography]{revtex4-1} 
\pdfoutput=1

\def\be{\begin{equation}}
\def\ee{\end{equation}}
\def\bea{\begin{eqnarray}}
\def\eea{\end{eqnarray}}
\def\bml{\begin{subequations}}

\def\elea{\end{eqnarray}\end{subequations}}
\def\ba{\mathbf{a}}
\def\bb{\mathbf{b}}
\def\bx{\mathbf{x}}

\usepackage{graphicx}

\begin{document}

\title{Gravitational smoothing of kinks on cosmic string loops}

\author{Jeremy M. Wachter}
\email{Jeremy.Wachter@tufts.edu}
\author{Ken D. Olum}
\email{kdo@cosmos.phy.tufts.edu}
\affiliation{Institute of Cosmology, Department of Physics and Astronomy,\\ 
Tufts University, Medford, MA 02155, USA}

\begin{abstract}
We analyze the effect of gravitational back reaction on cosmic string loops with kinks, which is an important determinant of the shape, and thus the potential observability, of string loops which may exist in the universe today.  Kinks are not rounded off, but may be straightened out.  This means that back reaction will only cause loops with kinks to develop cusps after some potentially large fraction of their lifetimes. In some loops, symmetries prevent even this process, so that the loop evaporates in a self-similar fashion and the kinks are unchanged.  As an example, we discuss back-reaction on the rectangular Garfinkle-Vachaspati loop.
\end{abstract}


\maketitle

Cosmic strings are effectively one-dimensional objects which may have formed through brane inflation or symmetry breaking transitions in the early universe.  If cosmic strings exist, there is a network consisting of super-horizon strings and a distribution of loops.  For a review, see Ref.~\cite{Vilenkin:book}.  The long strings lose energy into loop production, and (if there are no couplings to other low-mass fields) the loops decay by emission of gravitational waves.

These gravitational waves have several important roles.  First, they allow string loops to dissipate, preventing them from redshifting like matter always, which would cause a ``string loop problem'' analogous to the monopole problem.  Second, back reaction from gravitational wave emission may affect the shape of the loops, so that old loops are different from ones which have been newly formed from the long string network \cite{Blanco-Pillado:2015ana}.  Finally, gravitational waves in the form of a stochastic background or bursts due to cusps on the string loop are the leading candidate for an observable signal that would allow us to discover a string network \cite{Damour:2000wa,Damour:2001bk,Damour:2004kw,Siemens:2006vk,Blanco-Pillado:2013qja,Sanidas:2012ee,Arzoumanian:2015liz}. We will concentrate here on the effect of back reaction on the loop shape. We work in units where $c=1$.

Before we consider back reaction, a loop that is small compared to the Hubble distance can be considered to evolve in flat space.  In that case, the evolution of the loop can be written
\be\label{eqn:Nambu}
\bx(t,\sigma) = \frac12\left[\ba(v)+\bb(u)\right]\,,
\ee
where $v=t-\sigma$, $u=t+\sigma$, and $\ba$ and $\bb$ are functions periodic in the loop length $L$ whose tangent vectors obey $|\ba'|=|\bb'|= 1$.  Back reaction can be understood as slow modification of the functions $\ba$ and $\bb$, providing we are in the regime where $G\mu\ll 1$, where $\mu$ is the string tension and linear mass density, and $G$ is Newton's constant.

Since $|\ba'|=|\bb'|= 1$, $\ba'$ and $\bb'$ can be considered paths on the Kibble-Turok sphere of unit vectors \cite{Kibble:1982cb,Turok:1984cn,Garfinkle:1987yw}.  In the rest frame of the loop, these paths are closed and each has its center of mass at the center of the sphere.  If the string is smooth, these paths will generically (though not always \cite{Garfinkle:1987yw}) intersect. Such a point is called a cusp.  The string doubles back on itself there and moves (formally) at the speed of light.  Cusps lead to emission of bursts of gravitational waves, which may be detectable \cite{Damour:2000wa,Damour:2001bk,Damour:2004kw,Siemens:2006vk}.

However, the reconnection of strings during their evolution causes discontinuities in the paths of $\ba'$ and $\bb'$, and thus sudden changes in the direction of the strings, called kinks.  Kinks allow $\ba'$ and $\bb'$ to jump over each other and avoid forming a cusp, but lead to their own patterns of gravitational wave emission \cite{Garfinkle:1987yw}.

Simulations show that loops just formed from the long string network essentially never have cusps \cite{Blanco-Pillado:2015ana}.  Instead they have several large-angle kinks.  But most loops which exist at any given time have lost a very significant fraction of their length to gravitational radiation and thus may have very different shapes from those they had at formation.  If gravitational back reaction rounds off the kinks, then the paths of $\ba'$ and $\bb'$ will cross, producing cusps.  Generally there will be two cusps per oscillation \cite{Blanco-Pillado:2015ana}.

In certain systems, the effect of gravitational back reaction is easy to understand.  Consider, for example, a straight, static string with small wiggles of various wavelengths.  In this case one can straightforwardly compute the gravitational power from wiggles of some frequency $\omega$ and energy density $E$ interacting with other wiggles going the opposite direction \cite{Hindmarsh:1990xi,Siemens:2001dx}.  This power should come from a decrease in the energy of the wiggles in question, and so one arrives at the damping rate of these wiggles, $dE/dt\propto -E c\omega$, where $c$ is a constant depending on the opposite-direction wiggles (as long as the wiggles going in the two directions have similar wavelengths \cite{Siemens:2001dx}). After some interval $t$, $E$ will decrease by $e^{-t c \omega}$.

If this process were linear, one could argue that when there is a superposition of waves of different frequencies, the short waves are damped more rapidly in proportion to their wavelength.  Then the fate of any pattern of excitations on the string could be found by Fourier analysis, and we would conclude that effect of back reaction on the position of a string would be to convolve it with a Lorentzian, because the Lorentzian is the function whose Fourier transform declines exponentially with frequency.

Indeed, this was the procedure used in Ref.~\cite{Blanco-Pillado:2015ana} to produce a toy model of the effects of back reaction.  Convolution rounds off kinks, so the result was that cusps were generated, as described above.  However, radiation is not a linear process, and this model is qualitatively incorrect, as we will describe below.

Gravitational back reaction on string loops was studied numerically by Quashnock and Spergel \cite{Quashnock:1990wv}.  However their simulations were limited by the computer power available at the time, and so the detailed fate of the kinks is not clear. Ref.~\cite{Quashnock:1990wv} says that ``the kink angles are opened'', but the same results have been interpreted by other authors (e.g., Ref.~\cite{Vilenkin:book}) to show that the kinks are rounded off.

To determine whether rounding off, straightening out, or some other model is correct, we will analyze the general properties of back reaction.  We will consider strings with $G\mu\ll 1$, so we can work in linearized gravity.  Starting from the Minkowski metric $\eta_{\alpha\beta}$, the perturbed metric due to the string loop can be written $g_{\alpha\beta}= \eta_{\alpha\beta} + h_{\alpha\beta}$ with
\be\label{eqn:h0}
h_{\alpha\beta}(x)= -16G\pi\int d^4x' \left[T_{\alpha\beta}(x')-(1/2)\eta_{\alpha\beta}T^\gamma_\gamma(x')\right] D_r(x-x')\,,
\ee
where $T_{\alpha\beta}(x')$ is the stress-energy tensor of the string and $D_r$ is the retarded Green's function.

The curvature of spacetime leads to a change in the motion of the string, given by \cite{Quashnock:1990wv}
\be\label{eqn:force}
\frac{d^2x^\lambda}{du\,dv} = -\Gamma^\lambda_{\alpha\beta} \frac{dx^\alpha}{du} \frac{dx^\beta}{dv}\,,
\ee
where $\Gamma^\lambda_{\alpha\beta}$ is the Christoffel symbol.  Thus after some time, the function $\ba'(v) = 2d\bx/dv $ will change by the quantity \cite{Quashnock:1990wv},
\be
\Delta a'^\lambda = -2\int du\, \Gamma^\lambda_{\alpha\beta} \frac{dx^\alpha}{du} \frac{dx^\beta}{dv}\,,
\ee
and similarly for $\bb' (u)$.

How does $h_{\alpha\beta}(x)$ depend on the observation point $x$?
Equation~(\ref{eqn:h0}) is an integral over the intersection of the
past light cone of $x$ with the world sheet of the string.  If the
world sheet is smooth, moving $x$ changes this intersection in a
smooth fashion, so $h_{\alpha\beta}(x)$ changes smoothly.  Kinks in
the world sheet are lightlike lines where the world sheet has a slope
discontinuity.  Generically, the backward light cone from $x$
intersects such lines in isolated points.  Moving $x$ moves the
intersection points without changing their nature, and thus
$h_{\alpha\beta}(x)$ is still smooth.

When the backward light cone crosses the point where two kinks pass
through each other, this no longer applies.  The two kinks divide the
world sheet into four regions.  The backward light cone of $x$
intersects three of these, and moving $x$ causes the intersection with
one region to gradually decrease to zero and then the intersection
with the fourth region to increase from zero.  Thus there is a jump in
the first derivative of $h_{\alpha\beta}(x)$ at such points.  The
Christoffel symbol there has a discontinuity but remains bounded.

Now consider the effect of Eq.~(\ref{eqn:force}) on the string very
near a kink.  As we approach the kink, $dx/du$ and $dx/dv$ have
constant limits and $\Gamma^\lambda_{\alpha\beta}$ is bounded.  Thus
the force which changes the direction of the string approaches a
constant as we approach the kink tip, so the original direction is
perturbed in a uniform way.  The perturbed string near the kink may
point in a different direction, but it remains straight.  Thus kinks
are not rounded off.  Rounding off would require that after even a
small number of oscillations, the string very near the kink would be
substantially modified, which would be possible only if the force
diverged as one approached the kink, which is not the case.

As we cross over to the other side of the kink,
$\Gamma^\lambda_{\alpha\beta}$ has no sudden change, but $dx/du$ or
$dx/dv$ changes discontinuously to produce a sharp kink.  Thus the
force on the other side of the kink may be different.  This means that
the two sides may be turned in different directions and so the angle
of the kink may change.

If the kinks were rounded off, the paths of $\ba'$ and $\bb'$ would
become continuous, which would almost always lead to cusps. Opening
the kink angle makes cusps more likely, as the curvature introduced by
back reaction spreads out the unit vectors on the Kibble-Turok sphere
away from their discontinuous jumps, and so their paths may
overlap. But it is still possible for $\ba'$ to jump over $\bb'$, or
vice versa, so cusps are by no means certain.

One might argue that cusps become inevitable once back reaction
completely straightens kinks, but this is not guaranteed to
happen. Consider a kink due to a discontinuity in $A'$ with a very
shallow angle $\zeta\ll1$. This kink will change its angle in a way
that depends on the difference in force across the kink. There will be
a factor $G\mu$ from $h_{\alpha\beta}(x)$, a term like $1/L$ due to
the derivatives of $h_{\alpha\beta}(x)$ found in the Christoffel
symbol, and a dimensionless prefactor which depends on the contraction
of null vectors with the Christoffel symbol. Under the small-angle
approximation, the difference in null vectors will be linear in
$\zeta$, and so altogether we have
\be\label{eqn:dzdt}
\frac{d\zeta}{dt}=-\frac{\Gamma_\zeta G\mu}{L}\zeta\,.
\ee
The total radiation power of a loop is usually written $\Gamma G \mu^2$, with $\Gamma \sim 50$ depending on the loop shape.  As a result, the length of the loop decreases as $L(t)=L_0-\Gamma G\mu t$. If we put this into Eq.~(\ref{eqn:dzdt}) and solve the resulting differential equation for initial kink angle $\zeta_0$, we find
\be\label{eqn:small-angle}
\zeta(t)=\left(1-\frac{\Gamma G\mu t}{L_0}\right)^{\Gamma_\zeta/\Gamma}\zeta_0\,,
\ee
so the kink angle depends on the fraction of the loop remaining raised to some power. When $\Gamma_\zeta\sim\Gamma$, small kink angles are opened at about the same rate as the loop dissipates. If a kink was preventing a cusp, such $\Gamma_\zeta$ will lead to the cusp appearing on average when the loop has half evaporated.  Larger $\Gamma_\zeta$ causes cusps to appear sooner, and $\Gamma_\zeta\ll\Gamma$ means that small kinks persist very late in the lifetime of the loop, greatly reducing the frequency of cusps.  Unfortunately, we do not know at this point the typical magnitude of $\Gamma_\zeta$.

We have therefore shown from general arguments two important features of how back reaction affects loops. First, kinks are opened up rather than being smoothed out, and so the appearance of cusps on a loop will be delayed. Second, very small kinks may persist all the way to the end of a loop's life. Both of these results serve to change the distribution of loops with cusps, and therefore the signals we expect from loops.

\begin{figure}
\includegraphics[width=2in]{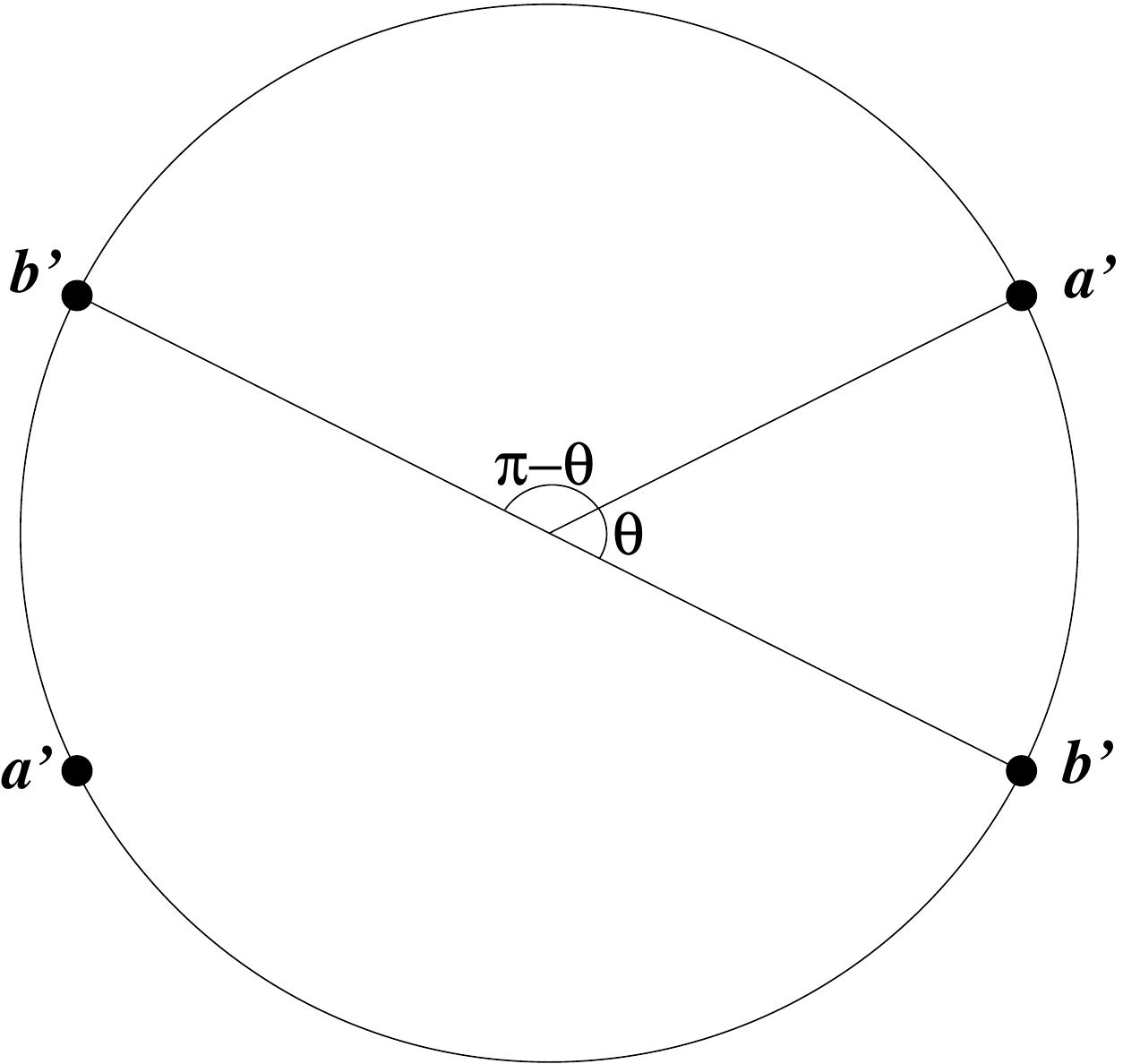}
\caption{The Kibble-Turok unit sphere for a rectangular loop.  Since the loop is planar, the sphere is just a circle, and each of $\ba'$ and $\bb'$ occupies 2 points.}
\label{fig:circle}
\end{figure}

\begin{figure}
\includegraphics[width=3in]{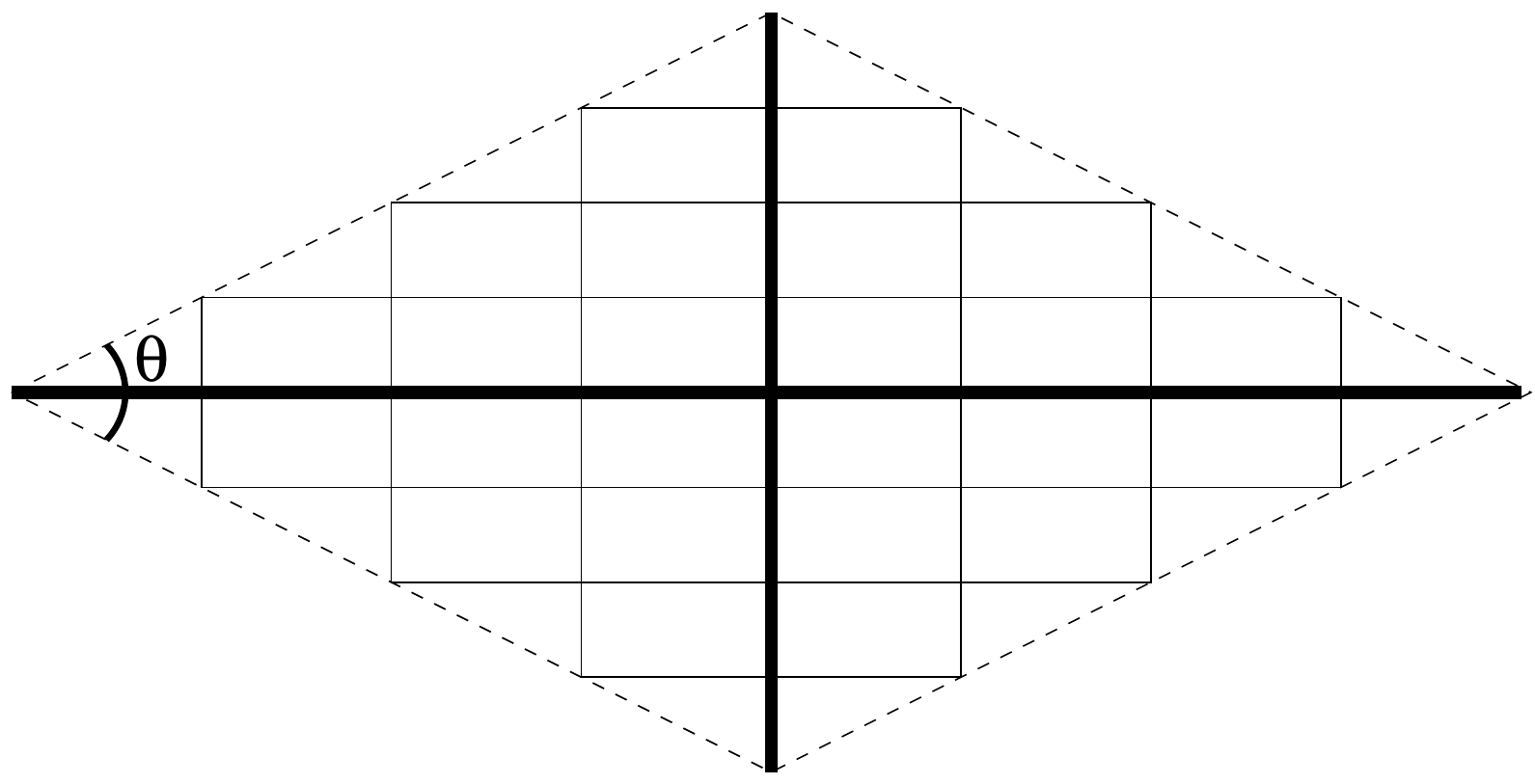}
\caption{A loop with straight segments and four kinks goes through a series of rectangular shapes, as shown here, including two double lines, shown bold. The angle between the directions of  $\ba'$ and $\bb'$ is $\theta$.}
\label{fig:rectangular}
\end{figure}

To demonstrate the processes we describe above, we consider the rectangular Garfinkle-Vachaspati loops~\cite{Garfinkle:1987yw,Garfinkle:1988yi}, which are simple enough that the first-order back reaction effect can be understood analytically \cite{Wachter:2016rwc}.  In such loops, the function $\ba'$ takes on two values only.  In the rest frame, these two vectors are the negatives of each other.  The function $\bb'$ is similar, so such a loop is specified up to overall length and spatial orientation by one number: the angle $\theta$ between the direction of one of the $\ba'$ and one of the $\bb'$ (and thus $\theta$ and $\pi-\theta$ specify the same loop).  The Kibble-Turok sphere for such a loop is shown in Fig.~\ref{fig:circle}.
The loop goes through a succession of rectangular shapes, with the limiting case of the rectangles being double lines, as shown in Fig.~\ref{fig:rectangular}. (Of course such intersections would destroy a real string, but here we are considering an infinitely thin string without self-interaction.)

What changes could back reaction produce in the shape of such a loop?
The loop lies in a plane, and reflection around this plane guarantees
that back reaction must leave the loop planar.  The functions $\ba$
and $\bb$, have angle-$\pi$ bends, and these could be reduced to less
sharp bends with curved segments in between.  This corresponds to
spreading out each point of $\ba'$ and $\bb'$ in
Fig.~\ref{fig:circle}, so that each curve moves slowly around the
circle for a while and then jumps by an angle less than $\pi$.

\begin{figure}
\includegraphics[width=6in]{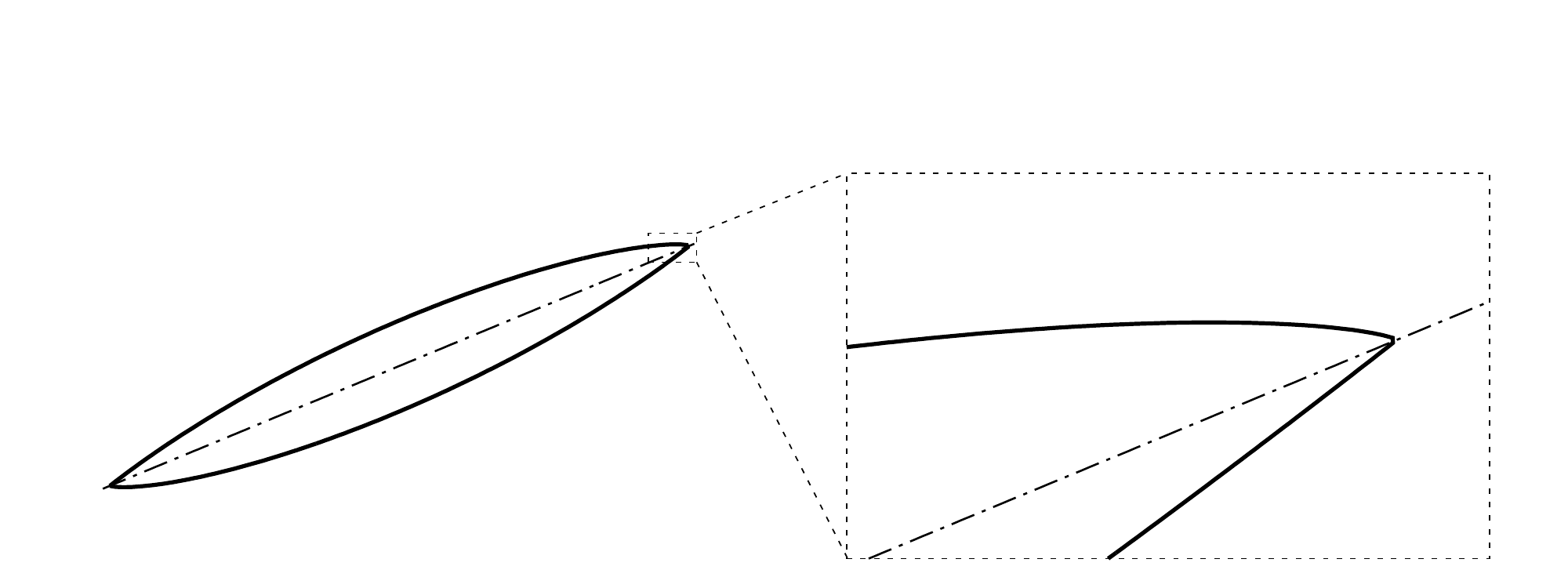}
\caption{The general effect of back reaction is to open out the kink angle, as shown here for the $\bb'$ of a Garfinkle-Vachaspati loop with $\theta=\pi/4$. Solid lines indicate the modified $\bb'$, while dot-dashed lines indicate the unmodified $\bb'$. Different forces acting before and after the kink lead to the segments there being turned by different amounts.}\label{fig:kinkmod}
\end{figure}

We analyzed these loops \cite{Wachter:2016rwc} using the formalism of
Eqs.~(\ref{eqn:h0},\ref{eqn:force}).  For general $\theta$, we found
that the kinks in $\ba$ and $\bb$ are opened out as shown by the
general argument above. Figure~\ref{fig:kinkmod} shows an example.

But in the case where $\theta = \pi/2$, there is an additional
symmetry.  The two directions of $\ba'$ lie equally at angle $\pi/2$
from the direction of $\bb'$, and vice versa.  Without any
calculation, one can see there is no way to decide whether $\ba'$ (or
$\bb'$) should be deformed to travel clockwise or counterclockwise
around the circle, and so it must remain unchanged.  Indeed, explicit
calculation \cite{Wachter:2016rwc} of this case shows that the loop
shrinks without changing shape.

Another example where kinks are not smoothed is the
Allen-Casper-Ottewill (ACO) loop \cite{Allen:1994bs} studied
extensively by Anderson
\cite{Anderson:2005qu,Anderson:2008wa,Anderson:2009rf}.  In this loop
$\ba$ is a circle, while $\bb$ is a line perpendicular to the plane of
$\ba$.  The loop is made of two segments of a helix, one right-handed
and one left-handed, and rotates rigidly.  In this case $\ba$ cannot
be smoothed because it is as smooth as possible already, and $\bb$
cannot be smoothed because of rotational invariance: any perturbation
to $\bb$ would require singling out a preferred direction.  Thus back
reaction cannot change the shape of this loop
\cite{Anderson:2005qu,Anderson:2008wa}.

The ACO loop can be extended to the case where $\ba$ wraps around the
circle more than once \cite{Allen:1994bs}, in which case the loop is
self-intersecting.  One can also have $\bb$ go back and forth several
times along the same line segment.  All such loops radiate without
change of shape.

In summary we find that back reaction does not round off kinks, but may change their angles, presumably straightening them out and moving some of the bending that was formerly at the kink into the segments between kinks.  Even this process does not always occur; in some cases symmetries prevent it and the loop shrinks due to evaporation, without changing its shape.  Both processes can be observed in the rectangular Garfinkle-Vachaspati loops, where the first-order effect of back reaction can be calculated analytically \cite{Wachter:2016rwc}.  But only the very simplest loops are amenable to analytic calculation, and even then only to first order, except when the loop shape remains the same.  Further progress will require numerical simulations in the style of Ref.~\cite{Quashnock:1990wv}, in particular to determine what fraction of loops develop cusps and at what point in their evolution, which is vital to determine the spectrum of gravitational waves produced by cosmic string loops.  This will be the subject of future work.

This work was supported in part by the National Science Foundation under grant number 1213930.  J.M.W. would like to acknowledge the support of the John F. Burlingame Graduate Fellowship.  We thank J. J. Blanco-Pillado for helpful conversations.

\bibliography{no-slac,letter}

\begin{thebibliography}{21}%
\makeatletter
\providecommand \@ifxundefined [1]{%
 \@ifx{#1\undefined}
}%
\providecommand \@ifnum [1]{%
 \ifnum #1\expandafter \@firstoftwo
 \else \expandafter \@secondoftwo
 \fi
}%
\providecommand \@ifx [1]{%
 \ifx #1\expandafter \@firstoftwo
 \else \expandafter \@secondoftwo
 \fi
}%
\providecommand \natexlab [1]{#1}%
\providecommand \enquote  [1]{``#1''}%
\providecommand \bibnamefont  [1]{#1}%
\providecommand \bibfnamefont [1]{#1}%
\providecommand \citenamefont [1]{#1}%
\providecommand \href@noop [0]{\@secondoftwo}%
\providecommand \href [0]{\begingroup \@sanitize@url \@href}%
\providecommand \@href[1]{\@@startlink{#1}\@@href}%
\providecommand \@@href[1]{\endgroup#1\@@endlink}%
\providecommand \@sanitize@url [0]{\catcode `\\12\catcode `\$12\catcode
  `\&12\catcode `\#12\catcode `\^12\catcode `\_12\catcode `\%12\relax}%
\providecommand \@@startlink[1]{}%
\providecommand \@@endlink[0]{}%
\providecommand \url  [0]{\begingroup\@sanitize@url \@url }%
\providecommand \@url [1]{\endgroup\@href {#1}{\urlprefix }}%
\providecommand \urlprefix  [0]{URL }%
\providecommand \Eprint [0]{\href }%
\providecommand \doibase [0]{http://dx.doi.org/}%
\providecommand \selectlanguage [0]{\@gobble}%
\providecommand \bibinfo  [0]{\@secondoftwo}%
\providecommand \bibfield  [0]{\@secondoftwo}%
\providecommand \translation [1]{[#1]}%
\providecommand \BibitemOpen [0]{}%
\providecommand \bibitemStop [0]{}%
\providecommand \bibitemNoStop [0]{.\EOS\space}%
\providecommand \EOS [0]{\spacefactor3000\relax}%
\providecommand \BibitemShut  [1]{\csname bibitem#1\endcsname}%
\let\auto@bib@innerbib\@empty
\bibitem [{\citenamefont {Vilenkin}\ and\ \citenamefont
  {Shellard}(2000)}]{Vilenkin:book}%
  \BibitemOpen
  \bibfield  {author} {\bibinfo {author} {\bibfnamefont {A.}~\bibnamefont
  {Vilenkin}}\ and\ \bibinfo {author} {\bibfnamefont {E.~P.~S.}\ \bibnamefont
  {Shellard}},\ }\href@noop {} {\emph {\bibinfo {title} {Cosmic Strings and
  other Topological Defects}}}\ (\bibinfo  {publisher} {Cambridge University
  Press},\ \bibinfo {address} {Cambridge},\ \bibinfo {year} {2000})\BibitemShut
  {NoStop}%
\bibitem [{\citenamefont {Blanco-Pillado}\ \emph {et~al.}(2015)\citenamefont
  {Blanco-Pillado}, \citenamefont {Olum},\ and\ \citenamefont
  {Shlaer}}]{Blanco-Pillado:2015ana}%
  \BibitemOpen
  \bibfield  {author} {\bibinfo {author} {\bibfnamefont {Jose~J.}\ \bibnamefont
  {Blanco-Pillado}}, \bibinfo {author} {\bibfnamefont {Ken~D.}\ \bibnamefont
  {Olum}}, \ and\ \bibinfo {author} {\bibfnamefont {Benjamin}\ \bibnamefont
  {Shlaer}},\ }\bibfield  {title} {\enquote {\bibinfo {title} {{Cosmic string
  loop shapes}},}\ }\href {\doibase 10.1103/PhysRevD.92.063528} {\bibfield
  {journal} {\bibinfo  {journal} {Phys. Rev.}\ }\textbf {\bibinfo {volume}
  {D92}},\ \bibinfo {pages} {063528} (\bibinfo {year} {2015})},\ \Eprint
  {http://arxiv.org/abs/1508.02693} {arXiv:1508.02693 [astro-ph.CO]}
  \BibitemShut {NoStop}%
\bibitem [{\citenamefont {Damour}\ and\ \citenamefont
  {Vilenkin}(2000)}]{Damour:2000wa}%
  \BibitemOpen
  \bibfield  {author} {\bibinfo {author} {\bibfnamefont {Thibault}\
  \bibnamefont {Damour}}\ and\ \bibinfo {author} {\bibfnamefont {Alexander}\
  \bibnamefont {Vilenkin}},\ }\bibfield  {title} {\enquote {\bibinfo {title}
  {{Gravitational wave bursts from cosmic strings}},}\ }\href {\doibase
  10.1103/PhysRevLett.85.3761} {\bibfield  {journal} {\bibinfo  {journal}
  {Phys. Rev. Lett.}\ }\textbf {\bibinfo {volume} {85}},\ \bibinfo {pages}
  {3761--3764} (\bibinfo {year} {2000})},\ \Eprint
  {http://arxiv.org/abs/gr-qc/0004075} {arXiv:gr-qc/0004075 [gr-qc]}
  \BibitemShut {NoStop}%
\bibitem [{\citenamefont {Damour}\ and\ \citenamefont
  {Vilenkin}(2001)}]{Damour:2001bk}%
  \BibitemOpen
  \bibfield  {author} {\bibinfo {author} {\bibfnamefont {Thibault}\
  \bibnamefont {Damour}}\ and\ \bibinfo {author} {\bibfnamefont {Alexander}\
  \bibnamefont {Vilenkin}},\ }\bibfield  {title} {\enquote {\bibinfo {title}
  {{Gravitational wave bursts from cusps and kinks on cosmic strings}},}\
  }\href {\doibase 10.1103/PhysRevD.64.064008} {\bibfield  {journal} {\bibinfo
  {journal} {Phys. Rev.}\ }\textbf {\bibinfo {volume} {D64}},\ \bibinfo {pages}
  {064008} (\bibinfo {year} {2001})},\ \Eprint
  {http://arxiv.org/abs/gr-qc/0104026} {arXiv:gr-qc/0104026 [gr-qc]}
  \BibitemShut {NoStop}%
\bibitem [{\citenamefont {Damour}\ and\ \citenamefont
  {Vilenkin}(2005)}]{Damour:2004kw}%
  \BibitemOpen
  \bibfield  {author} {\bibinfo {author} {\bibfnamefont {Thibault}\
  \bibnamefont {Damour}}\ and\ \bibinfo {author} {\bibfnamefont {Alexander}\
  \bibnamefont {Vilenkin}},\ }\bibfield  {title} {\enquote {\bibinfo {title}
  {{Gravitational radiation from cosmic (super)strings: Bursts, stochastic
  background, and observational windows}},}\ }\href {\doibase
  10.1103/PhysRevD.71.063510} {\bibfield  {journal} {\bibinfo  {journal} {Phys.
  Rev.}\ }\textbf {\bibinfo {volume} {D71}},\ \bibinfo {pages} {063510}
  (\bibinfo {year} {2005})},\ \Eprint {http://arxiv.org/abs/hep-th/0410222}
  {arXiv:hep-th/0410222 [hep-th]} \BibitemShut {NoStop}%
\bibitem [{\citenamefont {Siemens}\ \emph {et~al.}(2006)\citenamefont
  {Siemens}, \citenamefont {Creighton}, \citenamefont {Maor}, \citenamefont
  {Ray~Majumder}, \citenamefont {Cannon},\ and\ \citenamefont
  {Read}}]{Siemens:2006vk}%
  \BibitemOpen
  \bibfield  {author} {\bibinfo {author} {\bibfnamefont {Xavier}\ \bibnamefont
  {Siemens}}, \bibinfo {author} {\bibfnamefont {Jolien}\ \bibnamefont
  {Creighton}}, \bibinfo {author} {\bibfnamefont {Irit}\ \bibnamefont {Maor}},
  \bibinfo {author} {\bibfnamefont {Saikat}\ \bibnamefont {Ray~Majumder}},
  \bibinfo {author} {\bibfnamefont {Kipp}\ \bibnamefont {Cannon}}, \ and\
  \bibinfo {author} {\bibfnamefont {Jocelyn}\ \bibnamefont {Read}},\ }\bibfield
   {title} {\enquote {\bibinfo {title} {{Gravitational wave bursts from cosmic
  (super)strings: Quantitative analysis and constraints}},}\ }\href {\doibase
  10.1103/PhysRevD.73.105001} {\bibfield  {journal} {\bibinfo  {journal} {Phys.
  Rev.}\ }\textbf {\bibinfo {volume} {D73}},\ \bibinfo {pages} {105001}
  (\bibinfo {year} {2006})},\ \Eprint {http://arxiv.org/abs/gr-qc/0603115}
  {arXiv:gr-qc/0603115 [gr-qc]} \BibitemShut {NoStop}%
\bibitem [{\citenamefont {Blanco-Pillado}\ \emph {et~al.}(2014)\citenamefont
  {Blanco-Pillado}, \citenamefont {Olum},\ and\ \citenamefont
  {Shlaer}}]{Blanco-Pillado:2013qja}%
  \BibitemOpen
  \bibfield  {author} {\bibinfo {author} {\bibfnamefont {Jose~J.}\ \bibnamefont
  {Blanco-Pillado}}, \bibinfo {author} {\bibfnamefont {Ken~D.}\ \bibnamefont
  {Olum}}, \ and\ \bibinfo {author} {\bibfnamefont {Benjamin}\ \bibnamefont
  {Shlaer}},\ }\bibfield  {title} {\enquote {\bibinfo {title} {{The number of
  cosmic string loops}},}\ }\href {\doibase 10.1103/PhysRevD.89.023512}
  {\bibfield  {journal} {\bibinfo  {journal} {Phys. Rev.}\ }\textbf {\bibinfo
  {volume} {D89}},\ \bibinfo {pages} {023512} (\bibinfo {year} {2014})},\
  \Eprint {http://arxiv.org/abs/1309.6637} {arXiv:1309.6637 [astro-ph.CO]}
  \BibitemShut {NoStop}%
\bibitem [{\citenamefont {Sanidas}\ \emph {et~al.}(2012)\citenamefont
  {Sanidas}, \citenamefont {Battye},\ and\ \citenamefont
  {Stappers}}]{Sanidas:2012ee}%
  \BibitemOpen
  \bibfield  {author} {\bibinfo {author} {\bibfnamefont {S.~A.}\ \bibnamefont
  {Sanidas}}, \bibinfo {author} {\bibfnamefont {R.~A.}\ \bibnamefont {Battye}},
  \ and\ \bibinfo {author} {\bibfnamefont {B.~W.}\ \bibnamefont {Stappers}},\
  }\bibfield  {title} {\enquote {\bibinfo {title} {{Constraints on cosmic
  string tension imposed by the limit on the stochastic gravitational wave
  background from the European Pulsar Timing Array}},}\ }\href {\doibase
  10.1103/PhysRevD.85.122003} {\bibfield  {journal} {\bibinfo  {journal} {Phys.
  Rev.}\ }\textbf {\bibinfo {volume} {D85}},\ \bibinfo {pages} {122003}
  (\bibinfo {year} {2012})},\ \Eprint {http://arxiv.org/abs/1201.2419}
  {arXiv:1201.2419 [astro-ph.CO]} \BibitemShut {NoStop}%
\bibitem [{\citenamefont {Arzoumanian}\ \emph {et~al.}(2016)\citenamefont
  {Arzoumanian} \emph {et~al.}}]{Arzoumanian:2015liz}%
  \BibitemOpen
  \bibfield  {author} {\bibinfo {author} {\bibfnamefont {Z.}~\bibnamefont
  {Arzoumanian}} \emph {et~al.} (\bibinfo {collaboration} {NANOGrav}),\
  }\bibfield  {title} {\enquote {\bibinfo {title} {{The NANOGrav Nine-year Data
  Set: Limits on the Isotropic Stochastic Gravitational Wave Background}},}\
  }\href {\doibase 10.3847/0004-637X/821/1/13} {\bibfield  {journal} {\bibinfo
  {journal} {Astrophys. J.}\ }\textbf {\bibinfo {volume} {821}},\ \bibinfo
  {pages} {13} (\bibinfo {year} {2016})},\ \Eprint
  {http://arxiv.org/abs/1508.03024} {arXiv:1508.03024 [astro-ph.GA]}
  \BibitemShut {NoStop}%
\bibitem [{\citenamefont {Kibble}\ and\ \citenamefont
  {Turok}(1982)}]{Kibble:1982cb}%
  \BibitemOpen
  \bibfield  {author} {\bibinfo {author} {\bibfnamefont {T.~W.~B.}\
  \bibnamefont {Kibble}}\ and\ \bibinfo {author} {\bibfnamefont {Neil}\
  \bibnamefont {Turok}},\ }\bibfield  {title} {\enquote {\bibinfo {title}
  {{Selfintersection of Cosmic Strings}},}\ }\href {\doibase
  10.1016/0370-2693(82)90993-5} {\bibfield  {journal} {\bibinfo  {journal}
  {Phys. Lett.}\ }\textbf {\bibinfo {volume} {B116}},\ \bibinfo {pages}
  {141--143} (\bibinfo {year} {1982})}\BibitemShut {NoStop}%
\bibitem [{\citenamefont {Turok}(1984)}]{Turok:1984cn}%
  \BibitemOpen
  \bibfield  {author} {\bibinfo {author} {\bibfnamefont {Neil}\ \bibnamefont
  {Turok}},\ }\bibfield  {title} {\enquote {\bibinfo {title} {{Grand Unified
  Strings and Galaxy Formation}},}\ }\href {\doibase
  10.1016/0550-3213(84)90407-3} {\bibfield  {journal} {\bibinfo  {journal}
  {Nucl. Phys.}\ }\textbf {\bibinfo {volume} {B242}},\ \bibinfo {pages}
  {520--541} (\bibinfo {year} {1984})}\BibitemShut {NoStop}%
\bibitem [{\citenamefont {Garfinkle}\ and\ \citenamefont
  {Vachaspati}(1987)}]{Garfinkle:1987yw}%
  \BibitemOpen
  \bibfield  {author} {\bibinfo {author} {\bibfnamefont {David}\ \bibnamefont
  {Garfinkle}}\ and\ \bibinfo {author} {\bibfnamefont {Tanmay}\ \bibnamefont
  {Vachaspati}},\ }\bibfield  {title} {\enquote {\bibinfo {title} {{Radiation
  From Kinky, Cuspless Cosmic Loops}},}\ }\href {\doibase
  10.1103/PhysRevD.36.2229} {\bibfield  {journal} {\bibinfo  {journal} {Phys.
  Rev.}\ }\textbf {\bibinfo {volume} {D36}},\ \bibinfo {pages} {2229} (\bibinfo
  {year} {1987})}\BibitemShut {NoStop}%
\bibitem [{\citenamefont {Hindmarsh}(1990)}]{Hindmarsh:1990xi}%
  \BibitemOpen
  \bibfield  {author} {\bibinfo {author} {\bibfnamefont {Mark}\ \bibnamefont
  {Hindmarsh}},\ }\bibfield  {title} {\enquote {\bibinfo {title}
  {{Gravitational radiation from kinky infinite strings}},}\ }\href {\doibase
  10.1016/0370-2693(90)90226-V} {\bibfield  {journal} {\bibinfo  {journal}
  {Phys. Lett.}\ }\textbf {\bibinfo {volume} {B251}},\ \bibinfo {pages}
  {28--33} (\bibinfo {year} {1990})}\BibitemShut {NoStop}%
\bibitem [{\citenamefont {Siemens}\ and\ \citenamefont
  {Olum}(2001)}]{Siemens:2001dx}%
  \BibitemOpen
  \bibfield  {author} {\bibinfo {author} {\bibfnamefont {Xavier}\ \bibnamefont
  {Siemens}}\ and\ \bibinfo {author} {\bibfnamefont {Ken~D.}\ \bibnamefont
  {Olum}},\ }\bibfield  {title} {\enquote {\bibinfo {title} {{Gravitational
  radiation and the small-scale structure of cosmic strings}},}\ }\href
  {\doibase 10.1016/S0550-3213(01)00353-4, 10.1016/S0550-3213(02)00874-X}
  {\bibfield  {journal} {\bibinfo  {journal} {Nucl. Phys.}\ }\textbf {\bibinfo
  {volume} {B611}},\ \bibinfo {pages} {125--145} (\bibinfo {year} {2001})},\
  \bibinfo {note} {[Erratum: Nucl. Phys.B645,367(2002)]},\ \Eprint
  {http://arxiv.org/abs/gr-qc/0104085} {arXiv:gr-qc/0104085 [gr-qc]}
  \BibitemShut {NoStop}%
\bibitem [{\citenamefont {Quashnock}\ and\ \citenamefont
  {Spergel}(1990)}]{Quashnock:1990wv}%
  \BibitemOpen
  \bibfield  {author} {\bibinfo {author} {\bibfnamefont {Jean~M.}\ \bibnamefont
  {Quashnock}}\ and\ \bibinfo {author} {\bibfnamefont {David~N.}\ \bibnamefont
  {Spergel}},\ }\bibfield  {title} {\enquote {\bibinfo {title} {{Gravitational
  Selfinteractions of Cosmic Strings}},}\ }\href {\doibase
  10.1103/PhysRevD.42.2505} {\bibfield  {journal} {\bibinfo  {journal} {Phys.
  Rev.}\ }\textbf {\bibinfo {volume} {D42}},\ \bibinfo {pages} {2505--2520}
  (\bibinfo {year} {1990})}\BibitemShut {NoStop}%
\bibitem [{\citenamefont {Garfinkle}\ and\ \citenamefont
  {Vachaspati}(1988)}]{Garfinkle:1988yi}%
  \BibitemOpen
  \bibfield  {author} {\bibinfo {author} {\bibfnamefont {D.}~\bibnamefont
  {Garfinkle}}\ and\ \bibinfo {author} {\bibfnamefont {T.}~\bibnamefont
  {Vachaspati}},\ }\bibfield  {title} {\enquote {\bibinfo {title} {{Fields due
  to kinky, cuspless, cosmic loops}},}\ }\href {\doibase
  10.1103/PhysRevD.37.257} {\bibfield  {journal} {\bibinfo  {journal} {Phys.
  Rev.}\ }\textbf {\bibinfo {volume} {D37}},\ \bibinfo {pages} {257--262}
  (\bibinfo {year} {1988})}\BibitemShut {NoStop}%
\bibitem [{\citenamefont {Wachter}\ and\ \citenamefont
  {Olum}(2016)}]{Wachter:2016rwc}%
  \BibitemOpen
  \bibfield  {author} {\bibinfo {author} {\bibfnamefont {Jeremy~M.}\
  \bibnamefont {Wachter}}\ and\ \bibinfo {author} {\bibfnamefont {Ken~D.}\
  \bibnamefont {Olum}},\ }\bibfield  {title} {\enquote {\bibinfo {title}
  {{Gravitational back reaction on piecewise linear cosmic string loops}},}\
  }\href@noop {} {\  (\bibinfo {year} {2016})},\ \Eprint
  {http://arxiv.org/abs/1609.01685} {arXiv:1609.01685 [gr-qc]} \BibitemShut
  {NoStop}%
\bibitem [{\citenamefont {Allen}\ \emph {et~al.}(1994)\citenamefont {Allen},
  \citenamefont {Casper},\ and\ \citenamefont {Ottewill}}]{Allen:1994bs}%
  \BibitemOpen
  \bibfield  {author} {\bibinfo {author} {\bibfnamefont {Bruce}\ \bibnamefont
  {Allen}}, \bibinfo {author} {\bibfnamefont {Paul}\ \bibnamefont {Casper}}, \
  and\ \bibinfo {author} {\bibfnamefont {Adrian}\ \bibnamefont {Ottewill}},\
  }\bibfield  {title} {\enquote {\bibinfo {title} {{Analytic results for the
  gravitational radiation from a class of cosmic string loops}},}\ }\href
  {\doibase 10.1103/PhysRevD.50.3703} {\bibfield  {journal} {\bibinfo
  {journal} {Phys. Rev.}\ }\textbf {\bibinfo {volume} {D50}},\ \bibinfo {pages}
  {3703--3712} (\bibinfo {year} {1994})},\ \Eprint
  {http://arxiv.org/abs/gr-qc/9405037} {arXiv:gr-qc/9405037 [gr-qc]}
  \BibitemShut {NoStop}%
\bibitem [{\citenamefont {Anderson}(2005)}]{Anderson:2005qu}%
  \BibitemOpen
  \bibfield  {author} {\bibinfo {author} {\bibfnamefont {Malcolm~R.}\
  \bibnamefont {Anderson}},\ }\bibfield  {title} {\enquote {\bibinfo {title}
  {{Self-similar evaporation of a rigidly-rotating cosmic string loop}},}\
  }\href {\doibase 10.1088/0264-9381/22/13/002} {\bibfield  {journal} {\bibinfo
   {journal} {Class. Quant. Grav.}\ }\textbf {\bibinfo {volume} {22}},\
  \bibinfo {pages} {2539--2568} (\bibinfo {year} {2005})},\ \Eprint
  {http://arxiv.org/abs/gr-qc/0505160} {arXiv:gr-qc/0505160 [gr-qc]}
  \BibitemShut {NoStop}%
\bibitem [{\citenamefont {Anderson}(2009{\natexlab{a}})}]{Anderson:2008wa}%
  \BibitemOpen
  \bibfield  {author} {\bibinfo {author} {\bibfnamefont {Malcolm}\ \bibnamefont
  {Anderson}},\ }\bibfield  {title} {\enquote {\bibinfo {title} {{Continuous
  self-similar evaporation of a rotating cosmic string loop}},}\ }\href
  {\doibase 10.1088/0264-9381/26/2/025006} {\bibfield  {journal} {\bibinfo
  {journal} {Class. Quant. Grav.}\ }\textbf {\bibinfo {volume} {26}},\ \bibinfo
  {pages} {025006} (\bibinfo {year} {2009}{\natexlab{a}})},\ \Eprint
  {http://arxiv.org/abs/0812.2523} {arXiv:0812.2523 [gr-qc]} \BibitemShut
  {NoStop}%
\bibitem [{\citenamefont {Anderson}(2009{\natexlab{b}})}]{Anderson:2009rf}%
  \BibitemOpen
  \bibfield  {author} {\bibinfo {author} {\bibfnamefont {Malcolm}\ \bibnamefont
  {Anderson}},\ }\bibfield  {title} {\enquote {\bibinfo {title} {{Gravitational
  waveforms from the evaporating ACO cosmic string loop}},}\ }\href {\doibase
  10.1088/0264-9381/26/7/075018} {\bibfield  {journal} {\bibinfo  {journal}
  {Class. Quant. Grav.}\ }\textbf {\bibinfo {volume} {26}},\ \bibinfo {pages}
  {075018} (\bibinfo {year} {2009}{\natexlab{b}})},\ \Eprint
  {http://arxiv.org/abs/0903.4943} {arXiv:0903.4943 [gr-qc]} \BibitemShut
  {NoStop}%
\end{thebibliography}%

\end{document}